\documentclass[preprint,aps,12pt,showpacs,nofootinbib,tightenlines]{revtex4}
\usepackage{mathrsfs}
\usepackage{amsmath}
\usepackage{amssymb}
\usepackage{epsfig}
\usepackage{graphicx}
\usepackage{CJK}
\newcommand{\psl}{ P \hspace{-2.0truemm}/}
\newcommand{\esl}{ \epsilon \hspace{-1.5truemm}/ }

\def\be{\begin{eqnarray}}
\def\en{\end{eqnarray}}
\def\non{\nonumber\\}

\def\prd{{Phys. Rev. D}~}
\def\prl{{ Phys. Rev. Lett.}~}
\def\plb{{ Phys. Lett. B}~}

\def\epjc{{ Eur. Phys. J. C}~}
\newcommand{\acp}{{\cal A}_{CP}}
\begin{document}
\begin{CJK*}{GBK}{song}
\title{Analysis of $B \to a_1(1260)(b_1(1235))K^*$ decays in the perturbative QCD approach }
\author{Zhi-Qing Zhang
\\
{\small  \it Department of Physics, Henan University of
Technology, \\Zhengzhou, Henan 450052, P.R.China}}
\date{\today}
\begin{abstract}
Within the framework of perturbative QCD approach, we study the
charmless two-body decays $B\to a_1(1260)K^*, b_1(1235)K^*$. Using
the decays constants and the light-cone distribution amplitudes for
these mesons derived from the QCD sum rule method, we find the
following results: (a) Our predictions for the branching ratios are
consistent well with the QCDF results within errors, but much larger
than the naive factorization approach calculation values. (b) We
predict that the anomalous polarizations occurring in the decays
$B\to \phi K^*, \rho K^*$ also happen in the decays $B\to a_1K^*$,
while do not happen in the decays $B\to b_1K^*$. Here the
contributions from the annihilation diagrams play an important role
to explain the lager transverse polarizations in the decays $B\to
a_1K^*$, while they are not sensitive to the polarizations in decays
$B\to b_1K^*$. (c) Our predictions for the direct CP-asymmetries
agree well with the QCDF results within errors. The decays $\bar
B^0\to  b^{+}_1K^{*-}, B^-\to b^{0}_1K^{*-}$ have larger direct
CP-asymmetries, which could be measured by the present LHCb
experiments.
\end{abstract}

\vspace{1cm}

\maketitle


\section{Introduction}\label{intro}
In general, the mesons are classified in $J^{PC}$ multiplets. There
are two types of orbitally excited axial-vector mesons, namely,
$1^{++}$ and $1^{+-}$. The former includes $a_1(1260), f_1(1285),
f_1(1420)$, and $K_{1A}$, which compose the $^3P_1$ nonet, and the
latter includes $b_1(1235), h_{1}(1170), h_1(1380)$, and $K_{1B}$, which
compose the $^1P_1$ nonet. There is an important characteristic of these
axial-vector mesons, with the exception of  $a_1(1260)$ and $b_1(1235)$, that is, each different flavor state can mix
with another one, which comes from the other nonet meson or the same nonet.
There is not a mix between $a_1(1260)$ and $b_1(1235)$ because of
the opposite C parities. They do not also mix with others. So compared with other axial-vector mesons, these two mesons should have less
uncertainties regarding their inner structures.

Like the decay modes $B\to VV$, the charmless decays $B\to a_1(1260)K^*, b_1(1235)K^*$ also have three polarization states and so are expected to have rich physics.
In many $B\to VV$ decays, the information
on branching ratios and polarization fractions among various helicity amplitudes have been studied by many authors \cite{yli,ali,hwhuang,beneke}.
Through polarization studies, some underling helicity structures of the decay mechanism are proclaimed.
People find that the polarization fractions follow the naive counting rule, that is $f_L\sim 1-O(m^2_V/m^2_B), f_{N}\sim f_{T}\sim O(m_V^2/m_B^2)$
, where $f_{L, N, T}$ denote the longitudinal, parallel, and perpendicular polarization fractions, respectively, and $m_B(m_V)$ is the $B(V)$ meson mass.
But if the contributions from the factorizable emission
amplitudes are suppressed for some decay modes, this counting
rule might be modified to some extent even more dramatically
by other contributions. For example, many anomalous longitudinal polarization fractions in the decays $B\to \rho K^*, \phi K^*$ have been
measured by experiments, which are about $50\%$ \cite{hfag}, except that of the decay $B^-\to K^{*-}\rho^0$ with large value $(96^{+6}_{-16})\%$ \cite{hfag}
(the newer measurement is $(90\pm20)\%$) \cite{babar0}).
Whether a similar results also occurs in the decay modes $B\to a_1(1260)K^*, b_1(1235)K^*$
is worth researching. We know that $a_1(1260)$ has some similar behaviors as the vector meson,
so one can expect that there should exist some similar characteristics in the branching ratios and the
polarization fractions between the decays $B\to a_1(1260)K^*$ and $B\to \rho K^*$, where
$a_1(1260)$ and $\rho$  are scalar partners of each other, while this is not the case for $b_1(1235)$ because of its different characteristics
in the decay constant and light-cone distribution amplitude (LCDA) compared with those of $a_1(1260)$. For example, the longitude decay constant is very small for the charged
$b_1(1235)$ states and vanishes under the SU(3) limit. It is zero for the neutral $b^0_1(1235)$ state. While the transverse decay constant of $a_1(1260)$ vanishes under
the SU(3) limit. In the isospin limit, the chiral-odd (-even) LCDAs of meson $b_1(1235)$ are symmetric (antisymmetric) under the exchange of quark and antiquark momentum
fractions. It is just contrary to the symmetric behavior for $a_1(1260)$. In view of these differences, one can expect that there should exist very different
results between $B\to a_1(1260)K^*$ and $B\to b_1(1235)K^*$. On the
theoretical side, the decays $B\to a_1(1260)K^*, b_1(1235)K^*$ have been studied by Cheng and Yang in Ref. \cite{cheng}, where the branching ratios are very
different with those calculated by the naive factorization approach \cite{CMV}. To clarify such large differences is
another motivation of this work. On the experimental side, only the upper limits for some of the considered decays can be available \cite{babar2,babar3}.

In the following, $a_1(1260)$ and $b_1(1235)$ are denoted as $a_1$ and $b_1$ in some places for
convenience. The layout of this paper is as follows. In Sec.\ref{results}, we analyze these decay channels by using the PQCD
approach. The numerical results and the discussions are given in
Sec.\ref{numer}. The conclusions are presented in the final part.

\section{ the perturbative QCD  calculation} \label{results}
The PQCD approach has been proved been  an effective theory to
handle hadronic $B$ decays in many works
\cite{hwhuang,ali,PQCD,hnli0}. Because of taking into account the
transverse momentum of the valence quarks in the hadrons, one will
encounter double logarithm divergences when the soft and the
collinear momenta overlap. Fortunately, these large double logarithm
can be re-summed into the Sudakov factor \cite{hnli00}. There are
also another type of double logarithms which arise from the loop
corrections to the weak decay vertex. These double logarithms can
also be re-summed and resulted in the threshold factor, which
decreases faster than any other power of the momentum fraction in
the threshold region, which removes the endpoint singularity. This
factor is often parameterized into a simple form which is
independent on channels, twists and flavors \cite{hnli}. Certainly,
when the higher order diagrams only suffer from soft or collinear
infrared divergence, it is ease to cure by using the eikonal
approximation \cite{hnli2}. Controlling these kinds of divergences
reasonably makes the PQCD approach more self-consistent.

In the standard model, the related weak effective Hamiltonian
$H_{eff}$ mediating the $b\to s$ type transitions can be written as
\cite{buras96} \be {\cal H}_{eff} &=& \frac{G_{F}} {\sqrt{2}} \,
\left[\sum_{p=u,c}V_{pb} V_{ps}^* \left(C_1(\mu) O_1^p(\mu)
+ C_2(\mu) O_2^p(\mu) \right)-V_{tb} V_{ts}^*\sum_{i=3}^{10} C_{i}(\mu) \,O_i(\mu)
\right] . \label{eq:heff}
\en
Here the function $Q_i (i=1,...,10)$ is the local four-quark operator and $C_i$ is the corresponding Wilson coefficient.
$V_{p(t)b}, V_{p(t)s}$ are the CKM matrix elements. The standard four-quark operators are defined as:
\be
\begin{array}{llll}
O_1^{u} & = &  \bar s_\alpha\gamma^\mu L u_\beta\cdot \bar
u_\beta\gamma_\mu L b_\alpha\ , \quad\quad\quad\quad\quad\quad\;\; O_2^{u}  = \bar
s_\alpha\gamma^\mu L u_\alpha\cdot \bar
u_\beta\gamma_\mu L b_\beta\ , \\
O_3 & = & \bar s_\alpha\gamma^\mu L b_\alpha\cdot \sum_{q'}\bar
 q_\beta'\gamma_\mu L q_\beta'\ ,\quad\quad\quad\quad\;\;
O_4  =  \bar s_\alpha\gamma^\mu L b_\beta\cdot \sum_{q'}\bar
q_\beta'\gamma_\mu L q_\alpha'\ , \\
O_5 & = & \bar s_\alpha\gamma^\mu L b_\alpha\cdot \sum_{q'}\bar
q_\beta'\gamma_\mu R q_\beta'\ ,  \quad\quad\quad\quad\; O_6  =  \bar
s_\alpha\gamma^\mu L b_\beta\cdot \sum_{q'}\bar
q_\beta'\gamma_\mu R q_\alpha'\ , \\
O_7 & = & \frac{3}{2}\bar s_\alpha\gamma^\mu L b_\alpha\cdot
\sum_{q'}e_{q'}\bar q_\beta'\gamma_\mu R q_\beta'\ ,\quad\quad\quad   O_8  =
\frac{3}{2}\bar s_\alpha\gamma^\mu L b_\beta\cdot
\sum_{q'}e_{q'}\bar q_\beta'\gamma_\mu R q_\alpha'\ , \\
O_9 & = & \frac{3}{2}\bar s_\alpha\gamma^\mu L b_\alpha\cdot
\sum_{q'}e_{q'}\bar q_\beta'\gamma_\mu L q_\beta'\ ,  \quad\quad\quad O_{10}  =
 \frac{3}{2}\bar s_\alpha\gamma^\mu L b_\beta\cdot
\sum_{q'}e_{q'}\bar q_\beta'\gamma_\mu L q_\alpha'\ ,
\label{eq:operators}
\end{array}
\en
where $\alpha$ and $\beta$ are
the $SU(3)$ color indices; $L$ and $R$ are the left- and
right-handed projection operators with $L=(1 - \gamma_5)$, $R= (1 +
\gamma_5)$. The sum over $q'$ runs over the quark fields that are
active at the scale $\mu=O(m_b)$, i.e., $(q'\epsilon\{u,d,s,c,b\})$.
At leading order, there are eight types of single hard gluon exchange diagrams
contributing to our considered decays, dividing into the emission type diagrams and the annihilation
type diagrams, each type diagram including two factorizable ones and two nonfactorizable ones. Because of the limited space,we do not show these diagrams.

Combining the contributions from different diagrams, the total decay
amplitudes for these decays can be written as
\be \sqrt{2}{\cal M}_j(\bar
K^{*0}a^0_1)&=&\xi_u(F^{LL,j}_{eK^*}a_2+M^{LL,j}_{eK^*}C_2)-\xi_t\left[F^{LL,j}_{eK^*}\left(\frac{3C_{7}}{2}+
\frac{C_8}{2}+\frac{3C_{9}}{2}+\frac{C_{10}}{2}\right) \right.\non
&& \left.
-(F^{LL,j}_{ea_1}+F^{LL,j}_{aa_1})\left(a_4-\frac{a_{10}}{2}\right)
+M^{LL,j}_{eK^*}\frac{3C_{10}}{2} +M^{SP,j}_{eK^*}\frac{3C_{8}}{2}
\right.\non
&&\left.-(M^{LL,j}_{ea_1}+M^{LL,j}_{aa_1})\left(C_3-\frac{1}{2}C_9\right)
-(M^{LR,j}_{ea_1}+M^{LR,j}_{aa_1})\left(C_5-\frac{1}{2}C_7\right)\right.\non &&\left.-(F^{SP,j}_{ea_1}+F^{SP}_{aa_1})(a_6-\frac{1}{2}a_8)\right],\\\label{a10k0}
{\cal M}_j(\bar
K^{*0}a^-_1)&=&\xi_u\left[M^{LL,j}_{aa_1}C_1+F^{LL,J}_{aa_1}a_1\right]-\xi_t\left[F^{LL,j}_{ea_1}
\left(a_4-\frac{a_{10}}{2}\right)+F^{LL,j}_{aa_1}\left(a_4+a_{10}\right)
\right.\non
&&\left.+M^{LL,j}_{ea_1}\left(C_3-\frac{1}{2}C_9\right)+M^{LL,j}_{aa_1}\left(C_3+C_9\right)
+M^{LR,j}_{ea_1}\left(C_5-\frac{1}{2}C_7\right)
\right.\non &&\left.+M^{LR,j}_{aa_1}\left(C_5+C_7\right)+F^{SP,j}_{aa_1}(a_6+a_8)\right],\\
\sqrt{2}{\cal M}_j(\bar
K^{*-}a^0_1)&=&\xi_u\left[F^{LL,j}_{eK^*}a_2+M^{LL,j}_{eK^*}C_2+M^{LL,j}_{aa_1}C_1+F^{LL,j}_{aa_1}a_1\right]-\xi_t\left[
M^{LL,j}_{eK^*}\frac{3}{2}C_{10} \right.\non &&\left.
+M^{SP,j}_{eK^*}\frac{3}{2}C_{8}+M^{LL,j}_{aa_1}\left(C_3+C_9\right)+
M^{LR,j}_{aa_1}\left(C_5+C_7\right)\right.\non
&&\left.+F^{LL,j}_{aa_1}\left(a_4+a_{10}\right)
+F^{SP,j}_{aa_1}(a_6+a_8)\right],\\
{\cal M}_j(\bar
K^{*-}a^+_1)&=&\xi_u\left[F^{LL,j}_{ea_1}a_1+M^{LL,j}_{ea_1}C_1\right]-\xi_t\left[F^{LL,j}_{ea_1}\left(a_4+a_{10}\right)+
M^{LL,j}_{ea_1}\left(C_3+C_9\right)\right.\non
&&\left.+M^{LR,j}_{ea_1}\left(C_5+C_7\right)+M^{LL,j}_{aa_1}\left(C_3-\frac{1}{2}C_9\right)+
M^{LR,j}_{aa_1}\left(C_5-\frac{1}{2}C_7\right)\right.\non
&&\left.+F^{LL,j}_{aa_1}\left(a_4-\frac{1}{2}a_{10}\right)
+F^{SP,j}_{aa_1}\left(a_6-\frac{1}{2}a_8\right)\right],\label{a1zkf}
\en
here $F^{LL,j}_{ea_1}$ denotes the amplitudes of the factorizable
emission diagrams, where one can extract out the $B\to a_1$ transition form factor. If we replace the positions
of $a_1$ and $\bar K^*$ and will get the amplitudes
$F^{LL,j}_{eK^*}$ and $F^{SP,j}_{eK^*}$. As for the amplitudes of
non-factorizable emission diagrams,
$M^{LL,j}_{ea_1}$ and $M^{LR,j}_{ea_1}$ are relevant to the
considered decays. The amplitudes
$M^{LL,j}_{eK^*}$ and $M^{SP,j}_{eK^*}$ are obtained by exchanging
$a_1$ and $\bar K^*$ in these non-factorizable emission diagrams. It is similar to the
annihilation diagram amplitudes, where $F^{LL,j}_{aa_1}$ and
$F^{SP,j}_{aa_1}$ are for the factorizable ones, $M^{LL,j}_{aa_1}$
and $M^{LR,j}_{aa_1}$ are for the non-factorizable ones. It is
noticed that the upper labels $LL$, $LR$, and $SP$ denote the
$(V-A)(V-A)$, $(V-A)(V+A)$, and $(S-P)(S+P)$ currents, respectively,
and $j$ denotes three types of polarizations (one longitudinal and
two transverses), and named as $L, N, T$. Limitations of space prevent us from
giving the analytical expressions for these amplitudes. The
combinations of the Wilson coefficients are defined as usual: \be
a_{1}(\mu)&=&C_2(\mu)+\frac{C_1(\mu)}{3},
a_2(\mu)=C_1(\mu)+\frac{C_2(\mu)}{3},\quad\\
a_i(\mu)&=&C_i(\mu)+\frac{C_{i+1}(\mu)}{3},\quad
i=3,5,7,9,\\
a_i(\mu)&=&C_i(\mu)+\frac{C_{i-1}(\mu)}{3},\quad i=4, 6, 8,
10.\label{eq:aai} \en The amplitudes for those decays involving the
$b_1$ meson can be derived from the above expressions
Eq.(\ref{a10k0})-Eq.(\ref{a1zkf}) by substituting the $b_1$ meson
wave functions for $a_1$ ones.
\section{Numerical results and discussions} \label{numer}
For the wave function of the heavy B meson,
we take \cite{PQCD}
\be
\Phi_B(x,b)=
\frac{1}{\sqrt{2N_c}} (\psl_B +m_B) \gamma_5 \phi_B (x,b).
\label{bmeson}
\en
Here only the contribution of Lorentz structure $\phi_B (x,b)$ is taken into account, since the contribution
of the second Lorentz structure $\bar \phi_B$ is numerically small \cite{cdlu} and has been neglected. For the
distribution amplitude $\phi_B(x,b)$ in Eq.(\ref{bmeson}), we adopt the following model:
\be
\phi_B(x,b)=N_Bx^2(1-x)^2\exp[-\frac{M^2_Bx^2}{2\omega^2_b}-\frac{1}{2}(\omega_bb)^2],
\en
where $\omega_b$ is a free parameter, and taken as $\omega_b=0.4\pm0.04$ Gev in numerical calculations, and $N_B=91.745$
is the normalization factor for $\omega_b=0.4$. This is the same wave functions as in Ref.\cite{PQCD}, which is
a best fit for most of the measured hadronic B decays.

In these decays, both the longitudinal and the transverse polarizations are involved for the vector meson $K^*$.
Its distribution amplitudes are defined as
\be
\langle K^*(P, \epsilon^*_L)|\bar q_{2\beta}(z)q_{1\alpha}(0)|0\rangle&=&\frac{1}{\sqrt{2N_c}}\int^1_0dx \; e^{ixp\cdot z}
\left[m_{K^*}\esl^*_L\phi_{K^*}(x)+\esl^*_L \psl\phi_{K^*}^{t}(x)\right.\nonumber \\&&\left.
+m_{K^*}\phi^{s}_{K^*}(x)\right]_{\alpha\beta},\\
\langle K^*(P, \epsilon^*_T)|\bar q_{2\beta}(z)q_{1\alpha}(0)|0\rangle&=&\frac{1}{\sqrt{2N_c}}\int^1_0dx \; e^{ixp\cdot z}\left[m_{K^*}\esl^*_T\phi^v_{K^*}(x)+\esl^*_T \psl\phi_{K^*}^{T}(x)
\right.\non && \left.+m_{K^*}i\epsilon_{\mu\nu\rho\sigma}\gamma_5\gamma^\mu\epsilon^{*v}_Tn^\rho v^\sigma\phi^{a}_{K^*}(x)\right]_{\alpha\beta},
\en
where $n (v)$ is the unit vector having the same (opposite) direction with the moving of the vector meson and  $x$ is the momentum fraction of
$q_2$ quark. The upper (sub)leading twist wave functions can be parameterized as
\be
\phi_{K^*}(x)&=&\frac{f_{K^*}}{2\sqrt{2N_c}}\phi_\parallel(x), \phi^T_{K^*}(x)=\frac{f^T_{K^*}}{2\sqrt{2N_c}}\phi_\perp(x),\label{vaa}\\
\phi^t_{K^*}(x)&=&\frac{f^T_{K^*}}{2\sqrt{2N_c}}h^{(t)}_\parallel(x), \phi^s_{K^*}(x)=\frac{f^T_{K^*}}{2\sqrt{4N_c}}\frac{d}{dx}h^{(s)}_\parallel(x),\\
\phi^v_{K^*}(x)&=&\frac{f_{K^*}}{2\sqrt{2N_c}}g^{(v)}_\perp(x), \phi^a_{K^*}(x)=\frac{f_{K^*}}{8\sqrt{2N_c}}\frac{d}{dx}g^{(a)}_\perp(x), \label{vamp}
\en
where
\be
\phi_{\parallel,\perp}=6x(1-x)\left[1+3a^{\parallel,\perp}_{1K^*}t+3/2a^{\parallel,\perp}_{2K^*} (5t^2-1)\right],\\
h^{(t)}_\parallel(x)=3t^2,\quad\quad\quad\quad h^{(s)}_\parallel(x)=6x(1-x), \\
g^{(a)}_\perp(x)=6x(1-x),\quad\quad g^{(v)}_\perp(x)=3/4(1+t^2).
\en

For the distribution amplitudes of the axial-vectors $a_1(b_1)$, they have the same format as those of $K^*$ meson except the factor $i\gamma_5$ from the left hand:
\be
\langle A(P, \epsilon^*_L)|\bar q_{2\beta}(z)q_{1\alpha}(0)|0\rangle&=&\frac{i\gamma_5}{\sqrt{2N_c}}\int^1_0dx \; e^{ixp\cdot z}[m_A\esl^*_L\phi_A(x)+\esl^*_L \psl\phi_A^{t}(x)
+m_A\phi^{s}_A(x)]_{\alpha\beta},\non
\langle A(P, \epsilon^*_T)|\bar q_{2\beta}(z)q_{1\alpha}(0)|0\rangle&=&\frac{i\gamma_5}{\sqrt{2N_c}}\int^1_0dx \; e^{ixp\cdot z}\left[m_A\esl^*_T\phi^v_A(x)+\esl^*_T \psl\phi_A^{T}(x)
\right.\non && \left.+m_Ai\epsilon_{\mu\nu\rho\sigma}\gamma_5\gamma^\mu\epsilon^{*v}_Tn^\rho v^\sigma\phi^{a}_A(x)\right]_{\alpha\beta},
\en
where $A$ represents $a_1$ and $b_1$. Their (sub)leading twist wave functions have also the same parameter formats with those of $K^*$, which
can be gotten by replacing $K^*$ with $A$ in Eq.(\ref{vaa}$\sim$ \ref{vamp}). The corresponding functions $\phi(x), h(x), g(x)$ for axial-vector are written as
\be
\phi_{\parallel,\perp}&=&6x(1-x)\left[a^{\parallel,\perp}_0+3a^{\parallel,\perp}_1t+\frac{3a^{\parallel,\perp}_2}{2}(5t^2-1)\right],\\
h^{(t)}_\parallel(x)&=&3a^\perp_0t^2+\frac{3}{2}a^\perp_1t(3t^2-1),
h^{(s)}_\parallel(x)=6x(1-x)(a^\perp_0+a^\perp_1t),\\
g^{(a)}_\perp(x)&=&6x(1-x)(a^\parallel_0+a^\parallel_1t),
 g^{(v)}_\perp(x)=\frac{3}{4}a^\parallel_0(1+t^2)+\frac{3}{2}a^\parallel_1t^3,\label{t4}
\en
where the zeroth Gegenbauer moments $a^{\perp}_0(a_1)=a^{\parallel}_0(b_1)=0$ and $a^{\parallel}_0(a_1)=a^{\perp}_0(b_1)=1$. Here $t=2x-1$,
and other decay constants and Gegenbauer moments are listed in Table \ref{gegen}.
\begin{table}
\caption{Decay constants and Gegenbauer moments for $K^*$, $a_1$ and $b_1$ (in MeV). The values are taken at $\mu=1$ GeV.}
\begin{center}
\begin{tabular}{|c|c|c|c|}
\hline \hline  $f_{K^*}$ & $f^T_{K^*}$ &$f_{a_1}$&$f^T_{b_1}$\\
 $209\pm2$ &$165\pm9$&$238\pm10$ &$-180\pm8$\\
\hline
$a^\parallel_1(K^*)$&$a^\perp_1(K^*)$&$a^\parallel_2(K^*)$&$a^\perp_2(K^*)$\\
$0.03\pm0.02$&$0.04\pm0.03$&$0.11\pm0.09$&$0.10\pm0.08$\\
\hline
$a^\parallel_2(a_1(1260))$&$a^\perp_1(a_1(1260))$&$a^\parallel_1(b_1(1235))$&$a^\perp_2(b_1(1235))$\\
$-0.02\pm0.02$&$-1.04\pm0.34$&$-1.95\pm0.35$&$0.03\pm0.19$\\
\hline\hline
\end{tabular}\label{gegen}
\end{center}
\end{table}

The following input parameters are also used in our numerical calculations \cite{pdg10,ckmfit}:
\be
f_B&=&190 MeV, M_B=5.28 GeV, M_W=80.41 GeV,\\
\tau_{B^\pm}&=&1.638\times 10^{-12} s,\tau_{B^0}=1.525\times 10^{-12} s,\\
|V_{ub}|&=&3.89\times10^{-3}, |V_{tb}|=1.0,\gamma=(67.2\pm3.9)^\circ\\
|V_{us}|&=&0.2252,  |V_{ts}|=38.7\times10^{-3}.
\en

The matrix element ${\cal M}_{j}$ of the operators in the weak
Hamilitonian have been given in previous section, which are
rewritten as \be M_j&=&V_{ub}V^*_{us}T_{j}-V_{tb}V^*_{ts}P_{j}
=V_{ub}V^*_{us}T_{j}(1+z_je^{i(\gamma+\delta_j)}), \label{am} \en
where $\gamma$ is the Cabibbo-Kobayashi-Maskawa weak phase angle,
defined via $\gamma=arg[-\frac{V_{tb}V^*_{ts}}{V_{ub}V^*_{us}}]$.
$\delta_{j}$ is the relative strong phase between the tree and the
penguin amplitudes, which are denoted as "$T_{j}$" and "$P_{j}$",
respectively. The term $z_{j}$ describes the ratio of penguin to
tree contributions and is defined as \be
z_j=\left|\frac{V_{tb}V^*_{ts}}{V_{ub}V^*_{us}}\right|\left|\frac{P_j}{T_j}\right|.
\en In the same way, it is easy to write decay amplitude $\overline
{\cal M}_j$ for the corresponding conjugated decay mode: \be
\overline {\cal M}_j&=&V^*_{ub}V_{us}T_{j}-V^*_{tb}V_{ts}P_{j}
=V^*_{ub}V_{us}T_{j}(1+z_je^{i(-\gamma+\delta_j)}).\label{amcon}
\en So the CP-averaged branching ratio for each considered decay is
defined as
\be {\cal B}&=&\frac{G^2_F\tau_{B}}{32\hbar \pi m_B}(|{\cal
M}_j|^2+|\overline{\cal M}_j|^2)/2
\non &=&\frac{G^2_F\tau_{B}}{32\hbar
\pi
m_B}|V_{ub}V^*_{us}|^2\left[T^2_L(1+2z_L\cos\gamma\cos\delta_L+z_L^2)\right.\non
&&\left.
+2\sum_{i=N,T}T^2_i(1+2z_i\cos\gamma\cos\delta_i+z_i^2)\right].\label{brann}
\en Like the decays of $B$ to two vector mesons, there are also $3$
types of helicity amplitudes, so corresponding to $3$ types of $z_j$
and $\delta_j$, respectively. It is easy to see that the dependence
of decay width on $\delta$ and $\gamma$ is more complicated compared
with that for the decays of $B$ to two pseudoscalar mesons.

\begin{table}
\caption{ Branching ratios (in units of $10^{-6}$) for the decays
$B\to a_1(1260)K^*$ and $B\to b_1(1235)K^*$. In our
results, the errors for these entries correspond to the
uncertainties from $\omega_B$, the QCD scale $\Lambda^{(4)}_{QCD}$ and the threshold resummation parameter
$c$, respectively. For comparison, we also listed the results
predicted by QCDF approach \cite{cheng} and the naive factorization
approach \cite{CMV}.}
\begin{center}
\begin{tabular}{c|c|c|c}
\hline\hline   & This work  & \cite{cheng} & \cite{CMV}\\
\hline
$\bar B^0\to  a^{+}_1K^{*-}$ &$9.9^{+1.6+0.4+3.7}_{-1.1-0.6-3.7}$&$10.6^{+5.7+31.7}_{-4.0-8.1}$&$0.92$\\
$\bar B^0\to  a^{0}_1\bar K^{*0}$ &$7.1^{+1.5+0.4+3.1}_{-0.9-0.6-3.1}$&$4.2^{+2.8+15.5}_{-1.9-4.2} $&$0.64$\\
$B^-\to  a^{-}_1\bar K^{*0}$ &$10.8^{+2.0+0.7+4.6}_{-1.4-0.8-4.6}$&$11.2^{+6.1+31.9}_{-4.4-9.0} $&$0.51$\\
$B^-\to  a^{0}_1K^{*-}$ &$4.8^{+0.6+0.2+1.6}_{-0.5-0.3-1.6}$&$7.8^{+3.2+16.3}_{-2.5-4.3}$&$0.86$\\
\hline
$\bar B^0\to  b^{+}_1K^{*-}$ &$18.0^{+3.3+1.3+6.3}_{-2.6-2.3-6.3}$&$12.5^{+4.7+20.1}_{-3.7-9.0}$&$0.32$\\
$\bar B^0\to  b^{0}_1\bar K^{*0}$ &$9.6^{+2.1+1.0+3.8}_{-1.5-1.1-3.8}$&$6.4^{+2.4+8.8}_{-1.7-4.8} $&$0.15$\\
$B^-\to  b^{-}_1\bar K^{*0}$ &$23.0^{+4.5+2.3+8.4}_{-3.5-2.9-8.4}$&$12.8^{+5.0+20.1}_{-3.8-9.6} $&$0.18$\\
$B^-\to  b^{0}_1K^{*-}$ &$10.6^{+1.9+0.7+3.4}_{-1.5-1.4-3.4}$&$7.0^{+2.6+12.0}_{-2.0-4.8}$&$0.12$\\
\hline\hline
\end{tabular}\label{bran}
\end{center}
\end{table}

Using the input parameters as specified in
this section, it is easy to get the branching ratios for the
considered decays, which are listed in Table \ref{bran}, where the
first error comes from the uncertainty in the $B$ meson shape
parameter $\omega_b=0.40\pm0.04$ GeV, the second error is induced by
the hard scale-dependent varying from
$\Lambda^{(4)}_{QCD}=0.25\pm0.05$, and the last one is from the
threshold resummation parameter $c$ varying from $0.3$ to $0.4$.

In our predictions, the branching ratio of the decay $\bar B^0\to a^0_1\bar K^{*0}$ is larger than that of the decay $B^-\to a^0_1K^{*-}$, it is mainly
induced by the amplitudes of the factorizable emission diagrams, $F_{ea_1}$ and $F_{eK^*}$, have contrary interference effects between these two decays : constructive for the decay $a^0_1\bar K^{*0}$, destructive for the decay
$a^0_1K^{*-}$. So the decay $\bar B^0\to a^0_1\bar K^{*0}$ receives a larger real part for the penguin amplitudes. Though the decay
$B^-\to a^0_1K^{*-}$ has much larger contributions from tree ampllitudes, which are CKM suppressed and can not change the branching ratio too much.
In order to characterize the contribution from tree operators and the symmetry breaking effects between $B^-$ and $\bar B^0$ mesons, it is useful
to define the two ratios:
\be
R_1=\frac{{\cal B}(B^-\to a^-_1\bar K^{*0})}{{\cal B}(\bar B^0\to a^+_1K^{*-})}\times\frac{\tau_{\bar B^0}}{\tau_{B^-}},\quad
R_2=\frac{{\cal B}(B^-\to b^0_1K^{*-})}{{\cal B}(\bar B^0\to b^+_1K^{*-})}\times\frac{\tau_{\bar B^0}}{\tau_{B^-}}.
\en
If one neglects the tree operators and the electro-weak penguins, the ratios obey the following limits
\be
R_1=1, R_2=0.5.
\en
Here our predictions of these two ratios are $1.02$ and $0.55$, respectively. The results predicted by QCDF approach are $0.98$ and $0.52$, respectively.
If the future data for $R_1$ have large deviation from
our value, the contributions from electro-weak penguin operators might give an important affect, for the contribution from tree operators
can not change the branching ratio of $\bar B^0\to a^+_1K^{*-}$ too much. If the future data for $R_2$ have large deviation from our value,
some mechanism beyond factorization even from new physics might give an important affect, because the factorizaton formulae between
$\bar B^0\to b^+_1K^{*-}$ and $B^-\to b^0_1K^{*-}$ are exactly the same by considering
the neutral $b^0_1$ meson decay constant vanishing.

Compared with other results: From Table \ref{bran}, One can find that our predictions are consistent well with the QCDF results within (large) theoretical errors, while in stark
disagreement with the naive factorization approach, where the nonfactorizable effects are described by the effective number of colors $N^{eff}_c$. For some decays,
where the contributions from the emission diagrams are dominated or the branching ratios have a strong dependence on the correlative form factors, the
naive factorization approach can give a reasonable prediction, while for the decays, where the annihilation diagrams play an important role, this
approach would expose some disadvantages. On the experimental side, BarBar has been searched the decays $B\to a^-_1\bar K^{*0}, b_1K^*$ and set the upper limits on their
branching ratios ranging from $3.3$ to $8.0\times10^{-6}$ at the $90\%$ confidence level \cite{babar2,babar3}. Certainly, these upper limits are obtained by
assuming that ${\cal B}(a^{\pm}_1\to \pi^+\pi^-\pi^\pm)=
{\cal B}(a^{\pm}_1\to \pi^0\pi^0\pi^\pm)$ and ${\cal B}(a^{\pm}_1(b^\pm_1)\to \rho^{0}(\omega)\pi^{\pm})=1$. Furthermore, the background
signals may give an important effort on these upper limits, such as the background decay channel $B\to a_2\bar K^{*0}$ in studying of
the decay $B\to a_1\bar K^{*0}$.  In view of these disagreements, we strongly suggest
that the LHCb and the forthcoming Super-B experiments to accurately
measure these decays modes.

From Table \ref{polar}, we find that the polarization charactors for the decays $B\to a_1K^{*}$ and $B\to b_1K^{*}$ are very different: the transverse  polarization amplitudes have
almost equal values with (even a little stronger than) the longitudinal polarization ones for the former, while the longitudinal
polarization states are dominated for the latter. It seems that the anomalous polarizations occuring in decays $B\to \phi K^*, \rho K^*$
also happen in $B\to a_1K^*$ decays, while do not happen in $B\to b_1K^*$ decays. Here we also find that the contributions from the annihilation
diagrams are very important to the final polarization fractions for $B\to a_1K^{*}$ decays: If these contributions are neglected, the
longitudinal polarization fraction of the decay $B^-\to a_1^0K^{*-}$ becomes $98.8\%$, those of $\bar B^0\to a_1^+K^{*-}, a^0_1\bar K^{*0}$ increase to
about $90\%$, that of the decay $B^-\to a_1^-\bar K^{*0}$ changes from $50.3\%$ to $70.0\%$. While the longitudinal polarizations of decays
$B\to b_1K^{*}$ only have a very small decrease by neglecting the annihilation type contributions
except that of the decay $B^-\to b_1^-\bar K^{*0}$, which has a little large reduction, changing  from $96.2\%$ to $86\%$. In a word, the longitudinal polarizations
of decays $B\to b_1K^{*}$ are not very sensitive to the annihilation type contributions compared with those of $B\to a_1K^{*}$ decays.
\begin{table}
\caption{Longitudinal polarization fraction ($f_L$) and two transverse polarization fractions ($f_\parallel$, $f_\perp$) for decays
$B\to a_1(1260)K^*$ and $B\to b_1(1235)K^*$. In our results, the uncertainties come from $\omega_{B}$, the QCD scale $\Lambda^{(4)}_{QCD}$ and the threshold resummation
parameter $c$. The results of $f_L$ predicted by the QCDF approach are also displayed in parentheses for comparison.}
\begin{center}
\begin{tabular}{ccccccc|c}
\hline\hline   & $f_L(\%)$  &$f_\parallel(\%)$&$f_\perp(\%)$\\
\hline
$\bar B^0\to  a^{+}_1K^{*-}$&$48.9^{+5.1+7.4+4.9}_{-4.7-8.0-4.9}(37^{+39}_{-29})$&$26.1^{+2.5+3.8+2.6}_{-2.8-4.1-2.6}$&$25.0^{+2.2+3.8+2.3}_{-2.3-3.5-2.3}$\\
$\bar B^0\to  a^{0}_1\bar K^{*0}$ &$59.6^{+4.7+7.7+4.3}_{-4.9-7.8-4.3} (23^{+45}_{-19})$&$20.2^{+2.6+3.8+2.2}_{-2.5-3.8-2.2}$&$20.2^{+2.3+4.0+2.1}_{-2.2-3.5-2.1}$\\
$B^-\to  a^{-}_1\bar K^{*0}$&$50.3^{+5.1+8.6+5.0}_{-4.9-9.9-5.0} (37^{+48}_{-37})$&$24.1^{+2.6+5.0+2.5}_{-2.7-3.7-2.5}$&$25.6^{+2.3+5.0+2.5}_{-2.4-4.9-2.5}$\\
$B^-\to  a^{0}_1K^{*-}$ &$49.0^{+3.3+6.2+4.7}_{-4.3-6.2-4.7} (52^{+41}_{-42})$&$25.5^{+2.3+0.0+2.4}_{-2.5-2.5-2.4}$&$25.5^{+2.0+3.2+2.2}_{-2.2-3.7-2.2}$\\
\hline
$\bar B^0\to  b^{+}_1K^{*-}$&$95.9^{+0.1+1.1+0.0}_{-0.1-1.3-0.0}(82^{+18}_{-41})$&$1.1^{+0.2+0.4+0.2}_{-0.0-0.2-0.2}$&$3.0^{+0.0+0.9+0.2}_{-0.1-0.7-0.2}$\\
$\bar B^0\to  b^{0}_1\bar K^{*0}$ &$95.4^{+0.1+1.0+0.1}_{-0.1-1.4-0.1} (79^{+21}_{-74})$&$0.9^{+0.0+0.2+0.4}_{-0.0-0.2-0.4}$&$3.7^{+0.1+1.2+0.3}_{-0.1-0.8-0.3}$\\
$B^-\to  b^{-}_1\bar K^{*0}$&$96.2^{+0.0+0.9+0.1}_{-0.0-1.7-0.1} (79^{+21}_{-74})$&$1.0^{+0.0+0.3+0.3}_{-0.0-0.3-0.3}$&$2.8^{+0.0+0.9+0.2}_{-0.0-0.6-0.2}$\\
$B^-\to  b^{0}_1K^{*-}$ &$96.5^{+0.0+0.8+0.1}_{-0.1-1.3-0.1} (82^{+16}_{-26})$&$0.7^{+0.1+0.2+0.2}_{-0.0-0.1-0.2}$&$2.8^{+0.0+0.9+0.3}_{-0.0-0.6-0.3}$\\
\hline\hline
\end{tabular}\label{polar}
\end{center}
\end{table}
\begin{figure}[t,b]
\begin{center}
\includegraphics[scale=0.7]{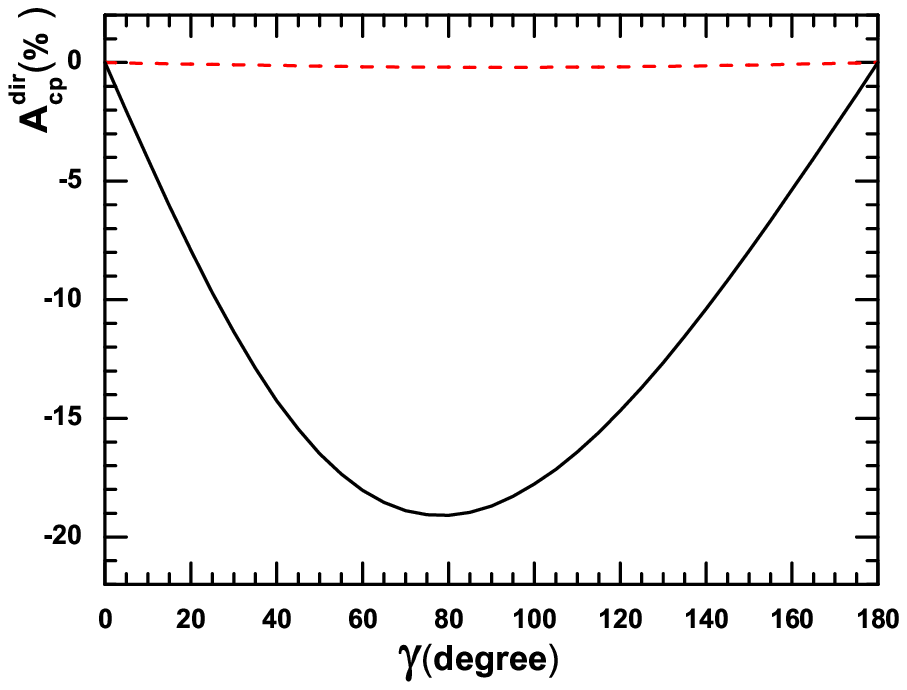}
\includegraphics[scale=0.7]{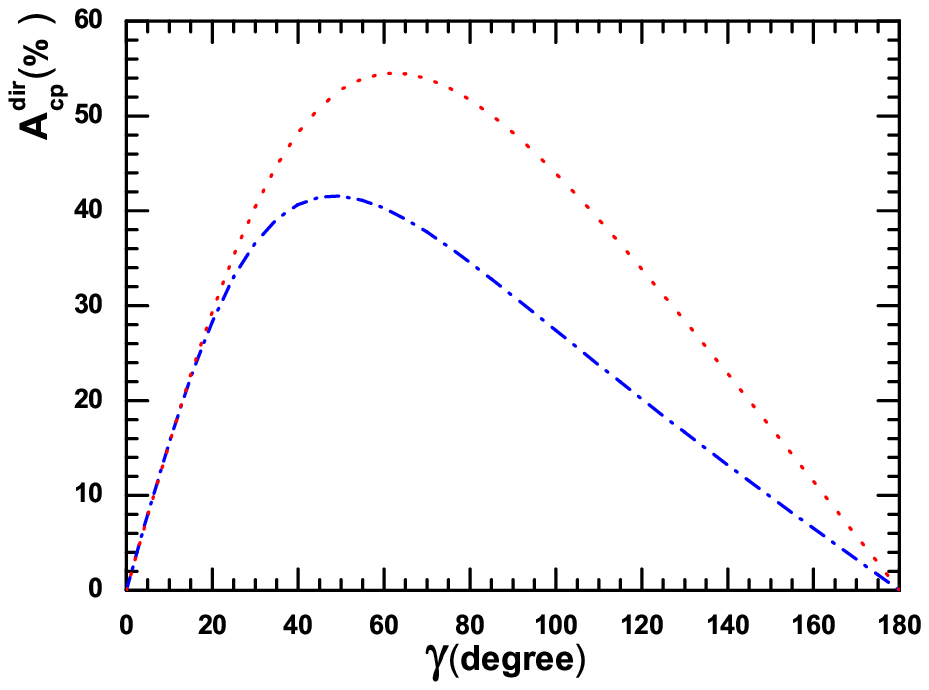}
\vspace{0.3cm} \caption{The dependence of the branching ratios on the
Cabibbo-Kobayashi-Maskawa angle $\gamma$. The left (right) panel is for the decays $B\to a_1(b_1)K^*$.
The dotted line represents the decays $B^-\to  a^{0}_1(b^0_1)K^{*-}$, the solid line represents the decays $\bar B^0\to  a^{0}_1(b^0_1)\bar K^{*0}$,
the dashed line is for the decays $B^-\to  a^{-}_1(b^{-}_1)\bar K^{*0}$, the dot-dashed line is for the decays $\bar B^0\to  a^{+}_1(b^{+}_1)K^{*-}$.}\label{fig2}
\end{center}
\end{figure}

Now we turn to the evaluations of the CP-violating asymmetries in PQCD approach. Here we only research the decays
$B\to b_1K^*$, where the
transverse polarization fractions are very small and range from $3.8$ to $5.2\%$.
It is easy to see that for these $b_1K^*$ decay modes, the contributions from the transverse polarizations are very small, so we neglected them in our calculations.
Using Eq.(\ref{am}) and Eq.(\ref{amcon}), one can get the expression for the direct CP-violating asymmetry:
\be
\acp^{dir}=\frac{ |\overline{\cal M}|^2-|{\cal M}|^2 }{
 |{\cal M}|^2+|\overline{\cal M}|^2} \nonumber  =\frac{2z_L\sin\alpha\sin\delta_L}
{(1+2z_L\cos\alpha\cos\delta_L+z_L^2) }\;.
\en

Using the input parameters and the wave functions as specified in this section, one can find the PQCD predictions (in units of $10^{-2}$) for the direct CP-violating asymmetries of the
considered decays:
\be
\acp^{dir}(\bar B^0\to  b^{+}_1K^{*-})&=&38.5^{+1.2+8.8+4.5}_{-1.7-7.4-4.5},\\
\acp^{dir}(B^-\to  b^{0}_1K^{*-})&=&54.3^{+0.9+7.8+4.4}_{-1.7-6.7-4.4},\\
\acp^{dir}(\bar B^0\to  b^{0}_1\bar K^{*0})&=&-18.7^{+2.0+0.7+1.8}_{-1.3-0.3-1.8},\\
\acp^{dir}(B^-\to  b^{-}_1\bar K^{*0})&=&-0.18^{+0.23+0.47+0.33}_{-0.28-0.00-0.33},
\en
where the errors are induced by the uncertainties of $B$ meson shape parameter $\omega_b=0.4\pm0.04$
, the hard scale-dependent varying from $\Lambda^{(4)}_{QCD}=0.25\pm0.05$, and the threshold resummation parameter $c$ varying from $0.3$
to $0.4$. In
Fig.\ref{fig2}, we show the Cabibbo-Kobayashi-Maskawa angle
$\gamma$ dependence of the direct CP-violating asymmetries of upper four decays. It is particularly noteworthy that our predictions about the
direct CP asymmetries of these decays are consistent well with the QCDF results \cite{kcyang} :
\be
\acp^{dir}(\bar B^0\to  b^{+}_1K^{*-})&=&(44^{+3}_{-58})\%,\\
\acp^{dir}(B^-\to  b^{0}_1K^{*-})&=&(60^{+6}_{-73})\%,\\
\acp^{dir}(\bar B^0\to  b^{0}_1\bar K^{*0})&=&(-17^{+21}_{-10})\%,\\
\acp^{dir}(B^-\to  b^{-}_1\bar K^{*0})&=&(2^{+0}_{-2})\%,
\en
where the error comes from the parameters $\rho_{A,H}$ and arbitrary phases $\phi_{A,H}$. These are phenomenological parameters to cure the endpoint
divergences in the amplitudes for the annihilation and hard spectator scattering diagrams.

\section{Conclusion}\label{summary}
In this paper, by using the decay constants and the light-cone distribution amplitudes
derived from QCD sum-rule method, we research  $B\to a_1K^*, b_1K^*$
decays in PQCD factorization approach and find that
\begin{itemize}
\item
Our predictions for the branching ratios are consistent well with the QCDF results within errors, but larger than
the naive factorization approach calculation values. On the experimental side, some primary upper limit values are inexplicable.
In view of these disagreements, we strongly suggest
that the LHCb and the forthcoming Super-B experiments can further accurately
measure these decays modes.
\item
The anomalous polarizations occuring  in decays $B\to \phi K^*, \rho K^*$
also happen in decays $B\to a_1K^*$, while do not happen in decays $B\to b_1K^*$. Here the contributions from the annihilation diagrams
play an important role to explain the lager transverse polarizations in decays $B\to a_1K^*$, while are not sensitive to the polarizations
in decays $B\to b_1K^*$.
\item
Our predictions for the direct CP-asymmetries agree well with the QCDF results within errors. The decays $\bar B^0\to  b^{+}_1K^{*-}, B^-\to  b^{0}_1K^{*-}$ have larger direct CP-asymmetries,
which could be measured by the present LHCb and the forthcoming Super-B experiments.
\end{itemize}

\section*{Acknowledgment}
This work is partly supported by the National Natural Science
Foundation of China under Grant No. 11147004, 11347030, by the Program of the Youthful Key Teachers in University of Henan Province
under Grant No. 001166, and by
Foundation of Henan University of Technology under Grant No.
2009BS038.

\end{CJK*}
\end{document}